\documentclass[aps,prc,superscriptaddress,twoside,twocolumn,%
nofootinbib,showpacs]{revtex4-1}
\usepackage{amsmath,amssymb}
\usepackage{graphicx}

\newcommand*{\fs}[1]{{#1\!\!\!/}}
\newcommand*{\hc}{\text{ H.\,c.}}
\newcommand*{\hQ}{\hat{Q}}

\newcommand*{\JT}{J_{\textsc{t}}}
\newcommand*{\tJ}{\tilde{J}}
\newcommand*{\tV}{\tilde{V}}
\newcommand*{\tB}{\tilde{B}}
\newcommand*{\GIP}{{\textsc{gip}}}
\newcommand*{\MEC}{{\textsc{mec}}}
\newcommand*{\IC}{{\text{int}}}
\begin{document}

\title{\boldmath Gauge-invariant formulation of $N N \to N N \gamma$}

\author{H. Haberzettl}
\email{helmut.haberzettl@gwu.edu}

\affiliation{\mbox{Center for Nuclear Studies, Department of Physics,
The George Washington University, Washington, DC 20052, U.S.A.}}

\author{K. Nakayama}
\email{nakayama@uga.edu}

\affiliation{Department of Physics and Astronomy, University of Georgia,
Athens, GA 30602, U.S.A.}
\affiliation{\mbox{Institut f{\"u}r Kernphysik and J\"ulich Center for Hadron Physics,
Forschungszentrum J{\"u}lich, 52425 J{\"u}lich, Germany}}

\date{8 November 2010}

\begin{abstract}
A complete, rigorous relativistic field-theory formulation of the
nucleon-nucleon ($NN$) bremsstrahlung reaction is presented. The resulting
amplitude is unitary as a matter of course and it is gauge invariant, i.e., it
satisfies a generalized Ward-Takahashi identity. The novel feature of this
approach is the consistent microscopic implementation of local gauge invariance
across all interaction mechanisms of the hadronic systems, thus serving as a
constraint for all subprocesses. The formalism is quite readily adapted to
approximations and thus can be applied even in cases where the microscopic
dynamical structure of the underlying interacting hadronic systems is either
not known in detail or too complex to be treated in detail. We point out how
the interaction currents resulting from the photon being attached to
nucleon-nucleon-meson vertices can be treated by phenomenological four-point
contact currents that preserve gauge invariance. In an advance application of
the present formalism [K. Nakayama and H. Haberzettl, Phys.\ Rev.\
C\,\textbf{80}, 051001 (2009)], such interaction currents were found to
contribute significantly in explaining experimental data. In addition, we
provide a scheme that permits --- through an introduction of phenomenological
five-point contact currents --- the approximate treatment of current
contributions resulting from pieces of the $NN$ interaction that cannot be
incorporated exactly. In each case, the approximation procedure ensures gauge
invariance of the entire bremsstrahlung amplitude. We also discuss the
necessary modifications when taking into account baryonic states other than the
nucleon $N$; in detail, we consider the $\Delta(1232)$ resonance by
incorporating the couplings of the $NN$ to the $N\Delta$ and $\Delta\Delta$
systems, and the $\gamma N\to \Delta$ transitions. We apply the formalism to
the 280-MeV bremsstrahlung data from TRIUMF [Phys.\ Rev.\ D\,\textbf{41}, 2689
(1990)] incorporating $\gamma N\Delta$ transition currents and find good
agreement with the data.
\end{abstract}



\maketitle

\section{Introduction}\label{sec:introduction}

The two-nucleon system is one of the simplest strongly interacting systems. The
study of the nucleon-nucleon ($NN$) bremsstrahlung reaction, therefore, offers
one of the most fundamental and direct avenues for understanding how the
electromagnetic field interacts with strongly interacting hadronic systems. In
the past, the $NN$ bremsstrahlung reaction had been applied extensively mainly
to learn about off-shell properties of the $NN$ interaction. It should be
clear, however, that off-shell effects are model-dependent and cannot be
measured, and therefore are meaningless quantities for the purpose of
comparison.

Even though the original motivation for investigating the $NN$ bremsstrahlung
reaction has fallen away, understanding the dynamics of such a fundamental
process, nevertheless, is of great importance from a general theoretical
perspective. This is highlighted by the fact that none of the past models of
$NN$ bremsstrahlung could describe the high-precision proton-proton
bremsstrahlung data from KVI~\cite{KVI02,KVI04} for coplanar geometries
involving small proton scattering angles. This was generally considered all the
more surprising since the irrelevance of off-shell effects was taken as
implicit proof positive that the coupling of a photon to the interacting
two-nucleon system was under control. The discrepancy between the KVI data and
the existing theoretical models, therefore, was quite unexpected. This
longstanding discrepancy of nearly a decade was resolved recently by the
present authors \cite{NH09} who put forward a novel approach to the $NN$
bremsstrahlung reaction that takes into account details of the photon coupling
to interacting systems that had previously been neglected. The study, in
particular, revealed the importance of accounting for the corresponding
interaction currents in a manner consistent with the gauge-invariance
constraint.

Another recent bremsstrahlung experiment concerns the hard bremsstrahlung
process $p + p \to pp(^1S_0) + \gamma$ measured for the first time by the
COSY-ANKE Collaboration \cite{ANKE}. In the absence of free systems of bound
diprotons, this process was considered as an alternative to the $\gamma +
pp(^1S_0) \to p + p$  process which complements the photo-disintegration of the
deuteron.  Here, the hardness of the bremsstrahlung is due to the fact that the
invariant mass of the two protons in the final state is constrained
experimentally to be less than 3 MeV above its minimum value of twice the
proton mass. In this kinematic regime, the two protons in the final state are
practically confined to the $^1S_0$ state and most of the available energy is
carried by the bremsstrahlung. Therefore, this kinematic regime is as far away
from the soft-photon limit as possible. The proton-proton ($pp$) hard
bremsstrahlung reaction has been also measured at CELSIUS-Uppsala \cite{JW09}.
In spite of extensive studies of the $NN$ bremsstrahlung reaction in the past,
no dedicated experiments of $pp$ hard bremsstrahlung with the diproton in the
final state had been available until these recent
measurements~\cite{ANKE,JW09}. Also, apart from the very recent study of
Ref.~\cite{NHUANG10}, the theoretical investigation of the $p + p \to pp(^1S_0)
+ \gamma$ reaction is virtually inexistent so far.

Apart from the intrinsic interest in the elementary $NN$ bremsstrahlung
process, the investigation of this reaction has also an immediate impact in the
area of heavy-ion physics. Indeed, di-lepton production in heavy-ion collisions
is used intensively as a probe of hadron dynamics in the nuclear medium. Due to
their weak interaction with hadrons, di-leptons are well suited to probe the
hadron dynamics in the dense region of heavy-ion collisions. The HADES
Collaboration, in particular, is currently engaged actively in the study of
di-electron production in heavy-ion as well as in elementary $NN$ collisions in
the 1-2 GeV/$u$ energy domain \cite{HADES}. In order to interpret the
experimental data in heavy-ion collisions, it is imperative to understand the
underlying basic elementary processes. Unfortunately, these basic elementary
processes are not yet fully under control. In fact, there is a number of
theoretical efforts to understand these basic reaction processes
\cite{KK06,SM09}. According to these studies, among the various competing
mechanisms, the $NN$ bremsstrahlung is one of the major mechanisms in producing
di-electrons in these reactions.

On the theoretical side, the majority of the existing models of $NN$
bremsstrahlung are potential models. They have been applied to the analyses of
experimental data (mostly in coplanar geometries) obtained up until the early
2000s, before the more recent experiments mentioned above were performed. Among
those models, the most recent and sophisticated ones that have been used in the
analysis of the high-precision KVI data \cite{KVI02,KVI04} are the microscopic
meson-exchange models of Refs.~\cite{MST97,CMSTT02,HNSA95}. There are also a
number of other microscopic model calculations throughout the 1990s
\cite{N89,KA93,J94,dJNHS94,JF95,dJNL95,EG96,KMS98,CST03}, which are dynamically
similar to Refs.~\cite{MST97,CMSTT02,HNSA95}, addressing a variety of issues in
the $NN$ bremsstrahlung process. All these models satisfy current conservation
(at least in the soft-photon approximation),\footnote{The current conservation
  of earlier models usually comes about because for $pp$ bremsstrahlung, there
  are no exchange currents for (uncharged) mesons and the four-point contact
  current discussed in Ref.~\cite{NH09} is absent for phenomenological
  meson-nucleon-nucleon form factors that depend only on the momentum of the
  exchanged meson. For $pn$ bremsstrahlung (see, e.g., Ref.~\cite{N89}) the
  meson-exchange currents are taken into account via Siegert's theorem which
  preserves current conservation in the soft-photon limit.}
but none of them obey the more general gauge-invariance condition in terms of
the generalized Ward-Takahashi identity (WTI) employed in Ref.~\cite{NH09} (and
explained in more detail in the present work). Quite recently, it was shown
formally~\cite{KBEHV09} how to maintain gauge invariance in a theory of
undressed non-relativistic nucleons if one introduces a finite cutoff in a
reference theory that is presumed to be already gauge invariant. When applied
to effective field theories, in particular, this implies that gauge invariance
can be maintained in such theories order-by-order in the expansion.

The purpose of the present paper is to present the complete, rigorous covariant
formulation of the $NN$ bremsstrahlung reaction whose successful advance
application was reported in Ref.~\cite{NH09}. The approach is based on a
relativistic field theory in which the photon is coupled in all possible ways
to the underlying two-nucleon $T$-matrix obtained from the corresponding
covariant Bethe-Salpeter-type $NN$ scattering equation. This formulation
follows the basic procedures of the field-theoretical approach of Haberzettl
\cite{H97} developed for pion photoproduction off the nucleon. The resulting
bremsstrahlung amplitude satisfies unitarity and gauge invariance as a matter
of course. The latter, in particular, is shown explicitly by deriving the
corresponding generalized WTI.

The formalism is quite readily adapted to approximations and thus can be
applied even in cases where the microscopic dynamical structure of the
underlying interacting hadronic systems is either not known in detail or too
complex to be treated in detail. As a case in point, we mention that the
success of the advance application~\cite{NH09} of the present formalism to the
KVI data~\cite{KVI02,KVI04} was due to the incorporation of phenomenological
four-point contact currents that preserve gauge invariance following the
approach of Haberzettl, Nakayama, and Krewald~\cite{HNK06} based on the
original ideas of Refs.~\cite{H97,HBMF98}. In addition, we provide a scheme
that permits the approximate treatment of current contributions resulting from
pieces of the $NN$ interaction that cannot be incorporated exactly. In each
case, the approximation procedure ensures gauge invariance of the entire
bremsstrahlung amplitude. We also discuss the necessary modifications when
taking into account baryonic states other than the nucleon $N$; in detail, we
consider the $\Delta(1232)$ resonance by incorporating the couplings of the
$NN$ to the $N\Delta$ and $\Delta\Delta$ systems, and the $\gamma N\to \Delta$
transitions.

In Sec.~\ref{sec:formalism}, we present the details of the full
four-dimensional relativistic formulation, including a proof of the gauge
invariance of the resulting bremsstrahlung amplitude. We also introduce the
necessary modifications for a covariant three-dimensional reduction and discuss
its implications for the description of the dynamics of the process. We point
out, in particular, that if one aims for a dynamically consistent
\emph{microscopic} description of all reaction mechanisms, one must implement
gauge invariance in terms of generalized Ward-Takahashi identities for each
subprocess --- mere global current conservation is not sufficient. We show how
one can preserve the gauge invariance of the amplitude even if some
interaction-current mechanisms --- both for hadronic three- and four-point
functions --- cannot be incorporated exactly. Furthermore, we discuss what
needs to be done to add additional baryonic degrees of freedom, in particular,
the coupling of $NN$, $N\Delta$, and $\Delta\Delta$ channels. In
Sec.~\ref{sec:Application}, we report on our results for the TRIUMF
data~\cite{TRIUMF90}. A summarizing assessment of the present work is given in
Sec.~\ref{sec:Summary}. Some technical details follow in the Appendix.

We emphasize that the present formalism is \emph{completely} general and
applies to proton-proton,  proton-neutron as well as neutron-neutron
bremsstrahlung processes. Furthermore, the photon can be either real or
virtual. The former corresponds to the usual $NN$ bremsstrahlung process while
the latter case is suited for applications in di-lepton production in both the
elementary $NN$ and heavy ion collisions.

\section{Formalism}\label{sec:formalism}

The bremsstrahlung current $B^\mu$ is obtained from the nucleon-nucleon
$T$-matrix by attaching an outgoing photon to all reaction mechanisms of $T$ in
all possible ways. To this end, we use here the gauge-derivative procedure
developed in Ref.~\cite{H97} in the context of pion photoproduction. This
procedure is formally equivalent to employing minimal substitution for the
connected part of the hadronic Green's function, and then taking the functional
derivative with respect to the electromagnetic four-potential $A^\mu$, in the
limit of vanishing $A^\mu$ (for details, see \cite{H97}). The current is then
obtained by removing the propagators of all external hadron legs from this
derivative in a Lehmann-Szymanzik-Zimmermann (LSZ) reduction
procedure~\cite{LSZ,WeinbergI}.

\subsection{Full four-dimensional formalism}\label{sec:4dimformalism}

The nucleon-nucleon $T$-matrix is determined by the corresponding
four-dimensional Bethe-Salpeter scattering equations,
\begin{equation}
  T=V+VG_0T
  \qquad\text{or}\qquad
  T= V+TG_0V~,
  \label{eq:NNLS}
\end{equation}
where $V$ is the $NN$ interaction given by the set of all two-nucleon
irreducible scattering mechanisms. $G_0$ describes the intermediate propagation
of two non-interacting nucleons, i.e., schematically we have
\begin{equation}
  G_0 =  [t_1\circ t_2]~,
  \label{eq:G0conv}
\end{equation}
where $t_i$ denotes the propagator of the individual nucleon $i$ and
``$\circ$'' stands for the convolution of the intermediate loop integration.

The bremsstrahlung current $\tB^\mu$ is obtained by evaluating the LSZ-type
equation
\begin{equation}
  \tB^\mu = -G_0^{-1} \left\{G_0 T G_0 \right\}^\mu G_0^{-1}
\end{equation}
where $\{\cdots\}^\mu$ denotes the gauge derivative~\cite{H97} taken here of
the connected hadronic $NN$ Green's function $G_0 T G_0$, with $\mu$ being the
Lorentz index of the current. Using then the product rule
$\{YX\}^\mu=Y\{X\}^\mu+\{Y\}^\mu X$ for an (ordered) product $YX$ of a two-step
sequence of hadronic reaction mechanisms described by  operators $X$ (first
step) and $Y$ (second step), we employ Eq.~(\ref{eq:NNLS}) repeatedly to find
\begin{equation}
  \tB^\mu
  =(1+TG_0) \left(d^\mu + V^\mu\right)(G_0 T +1) -d^\mu~,
  \label{eq:Bmu0}
\end{equation}
where $d^\mu$ defined by
\begin{equation}
d^\mu =- G_0^{-1}\{G_0\}^\mu G_0^{-1}
\label{eq:G0current0}
\end{equation}
subsumes the one-body current contributions from the individual nucleons and
\begin{equation}
  V^\mu = -\{V\}^\mu
\end{equation}
is the interaction current resulting from attaching the photon to any
\emph{internal} mechanisms of the $NN$ interaction. Details of $V^\mu$ will be
discussed below.

\begin{figure}[t!]\centering
\includegraphics[width=.5\columnwidth,clip=]{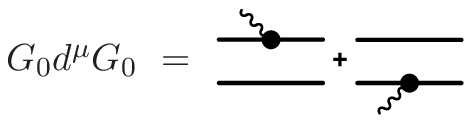}
  \caption{\label{fig:dmu}%
  Graphical representation of Eq.~(\ref{eq:G0current1}). Solid lines depict
  nucleons and wavy lines indicate the outgoing bremsstrahlung photon.}
\end{figure}

Explicitly, the photon contributions from the two-nucleon propagator are found as
\begin{equation}
  d^\mu = \Gamma_1^\mu (\delta_2 t_2^{-1}) + (\delta_1t_1^{-1})  \Gamma_2^\mu~,
  \label{eq:dgauge}
\end{equation}
where $\Gamma^\mu_i$ is the electromagnetic current operator of nucleon $i$;
$\delta_i$ denotes an implied $\delta$ function that makes the incoming and
outgoing momenta for the intermediate spectator nucleon $i$ the same. We thus
have
\begin{equation}
G_0 d^\mu G_0
 =\left[t_1 \Gamma_1^\mu t_1\circ t_2\right]
 +\left[ t_1 \circ t_2  \Gamma_2^\mu t_2\right]
\label{eq:G0current1}
\end{equation}
which is represented graphically in Fig.~\ref{fig:dmu}.

%
%

Note that Eq.~(\ref{eq:Bmu0}) --- apart from the subtraction by $d^\mu$ ---
possesses the structure of a distorted-wave Born approximation (DWBA), with the
factors $(G_0 T+1)$ and $(1+T G_0)$ supplying the M\o ller operators producing
the initial and final scattering states, respectively, distorted by the $NN$
interaction. The $d^\mu$ contribution by itself --- without any initial-state
or final-state $NN$ interactions --- is disconnected, as one sees clearly from
Fig.~\ref{fig:dmu}. The overall subtraction of $d^\mu$ in Eq.~(\ref{eq:Bmu0}),
therefore, is necessary to remove this (unphysical) disconnected structure from
$\tB^\mu$ and retain only connected physical contributions.

%
\begin{figure*}[t!]\centering
  \includegraphics[width=.7\textwidth,clip=]{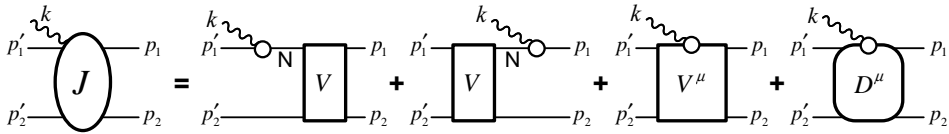}
  \caption{\label{fig:Jgeneric}%
  Basic production current $J^\mu$ of Eq.~(\ref{eq:JmuModified}) for $NN\to NN
  \gamma$. Time proceeds from right to left. External legs are labeled by the
  four-momenta of the respective particles. Boxes labeled $V$ subsume
  \emph{all} two-nucleon   irreducible contributions to the $NN$ interaction
  that drive the Bethe-Salpeter equation (\ref{eq:NNLS}). The interaction
  current $V^\mu$ contains all mechanisms where the photon emerges from
  \emph{within} the interaction $V$ (i.e., any current mechanism not associated
  with an external leg of the interaction $V$). The last diagram, $D^\mu$, as
  given in Eq.~(\ref{eq:DmuDefined}), subsumes all possible completely
  transverse contributions [cf.~Eq.~(\ref{eq:TmuStructure})], in addition to
  the subtraction that corrects the double counting arising from the first two
  contributions (see text). Diagrams where the photon emerges from the lower
  nucleon line are suppressed; the antisymmetrization of nucleons is implied.}
\end{figure*}
%

It is possible --- and indeed desirable for the following --- to equivalently
rewrite  Eq.~(\ref{eq:Bmu0}) to provide a full DWBA structure for $\tB^\mu$ of
the form
\begin{equation}
  \tB^\mu =(1+TG_0) \tJ^\mu(G_0 T +1)~,
  \label{eq:BmuJ}
\end{equation}
where
\begin{equation}
  \tJ^\mu = d^\mu G_0 V + VG_0 d^\mu +V^\mu -VG_0 d^\mu G_0 V
  \label{eq:Jmudefined}
\end{equation}
is the completely connected current that describes the bremsstrahlung reaction
in the absence of any hadronic initial-state or final-state interactions (with
the exception of the subtraction $VG_0 d^\mu G_0 V$; see below); we shall refer
to this tree-level-type current as the basic production current. The
equivalence of this form of $\tB^\mu$ to Eq.~(\ref{eq:Bmu0}) is easily seen by
repeated applications of Eq.~(\ref{eq:NNLS}). The $NN$ $T$-matrices appearing
to the right or left of the basic production current $J^\mu$ in (\ref{eq:BmuJ})
thus provide the initial-state interaction (ISI) or final-state
interaction(FSI), respectively, of the two nucleons external to the basic
production current $J^\mu$.

The subtraction term $VG_0 d^\mu G_0 V$ in $J^\mu$ removes here the double
counting of contributions $TG_0 d^\mu G_0 T$ to the full current $\tB^\mu$ that
arise from the two $d^\mu$ contributions in $\tJ^\mu$. As such, therefore, it
is not a dynamically independent contribution to $\tB^\mu$ and appears here
only because, for formal reasons, we wish to retain the DWBA form of $\tB^\mu$
in Eq.~(\ref{eq:BmuJ}). We shall see below, in Sec.~\ref{sec:3dimred}, where we
treat the covariant three-dimensional reduction of Eq.~(\ref{eq:BmuJ}), that
special considerations are needed for this subtraction term if one wants to
maintain gauge invariance for the reduced amplitude.

Equations (\ref{eq:BmuJ}) and (\ref{eq:Jmudefined}) are \emph{not} the complete
solution of the bremsstrahlung problem yet, because the gauge-derivative
procedure --- indeed any procedure based on minimal substitution --- cannot
produce current contributions that are completely transverse. Such
contributions must be added to the mechanisms obtained above for $\tB^\mu$.
Without lack of generality, we may do so by modifying the basic production
current $\tJ^\mu$ according to
\begin{equation}
 J^\mu = d^\mu G_0 V + VG_0 d^\mu +V^\mu + D^\mu
\label{eq:JmuModified}
\end{equation}
where
\begin{equation}
  D^\mu = T^\mu-VG_0 d^\mu G_0 V
  \label{eq:DmuDefined}
\end{equation}
contains the sum of all explicitly \emph{transverse} five-point currents
denoted by $T^\mu$, in addition to the subtraction current $VG_0 d^\mu G_0 V$.
In other words,
\begin{equation}
k_\mu T^\mu =0
\end{equation}
is true irrespective of whether the external nucleons are on-shell or not.  The
complete bremsstrahlung current then is given by
\begin{equation}
  B^\mu =(1+TG_0) J^\mu(G_0 T +1)~.
  \label{eq:BmuGeneric}
\end{equation}
We emphasize that this equation and Eq.~(\ref{eq:JmuModified}) provide an exact
generic description of the bremsstrahlung process off the $NN$ system. The
structure of the basic production current $J^\mu$ is depicted in
Fig.~\ref{fig:Jgeneric}.

%
\begin{figure}[t!]\centering
  \includegraphics[width=.3\columnwidth,clip=]{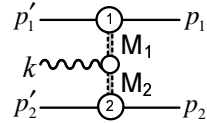}
  \caption{\label{fig:JMeson}%
  Generic example for meson transition currents $J^\mu_M$ containing a
  \emph{transverse} $\gamma M_1 M_2$ vertex, with $M_1\neq M_2$, occurring in
  $D^\mu$ of Eq.~(\ref{eq:DmuDefined}) via its transverse contribution $T^\mu$.
  The present numerical results were obtained incorporating
  $\gamma\rho\pi$ and $\gamma\omega\pi$ transition currents.}
\end{figure}
%

The detailed nature of the transverse contribution $T^\mu$ in
(\ref{eq:DmuDefined}) must be specified by the underlying interaction
Lagrangians. Examples are meson-transition currents $J^\mu_M$ as depicted in
Fig.~\ref{fig:JMeson} and $\gamma N\Delta$-transition currents $J^\mu_\Delta$,
i.e.,
\begin{equation}
  T^\mu = J^\mu_M + J^\mu_\Delta+\cdots~.
  \label{eq:TmuStructure}
\end{equation}
The latter will be discussed below, in Sec.~\ref{sec:delta}, when we consider
$\Delta(1232)$ contributions in detail.

The specific details of any particular application, of course, depend on the
mechanisms taken into account in the $NN$ interaction $V$ that drives the
scattering process in the Bethe-Salpeter equation (\ref{eq:NNLS}). For driving
interactions based on single-meson exchanges, the complete structure of $J^\mu$
is discussed below, in Sec.~\ref{subsec:singlemeson}. In Sec.~\ref{sec:delta},
as mentioned already, we also consider the structures arising from
$\Delta$ contributions that go beyond single-meson exchanges.

\subsection{Gauge invariance of the full four-dimensional amplitude}\label{sec:4dimgauge}

It should be clear that the procedure used for deriving the bremsstrahlung
current $B^\mu$ does produce a current that is gauge invariant as a matter of
course. Nevertheless, we will now explicitly prove gauge invariance because
this will guide us later, in Sec.~\ref{sec:3dimred}, in how to implement gauge
invariance when we calculate $B^\mu$ in a covariant three-dimensional
reduction.

To prove the gauge invariance of the current (\ref{eq:BmuGeneric}), we first
note that for an outgoing photon with four-momentum $k$, the four-divergence of
$d^\mu$ may be schematically written as
\begin{equation}
k_\mu d^\mu =  \hQ G_0^{-1}-G_0^{-1} \hQ ~,
\label{eq:divdmu}
\end{equation}
where $\hQ$ is short for
\begin{equation}
\hQ =\hQ_1 +\hQ_2~.
\label{eq:Qhatintro}
\end{equation}
The operator $\hQ_i$ describes the charge $Q_i$ of nucleon $i$ \emph{and} it
removes the four-momentum $k$ carried away by the photon from any subsequent
interaction of nucleon $i$ (appearing on the left of $\hQ_i$). This notation
allows one to keep track of the kinematics without explicit four-momentum
arguments, i.e., specifying initial momenta for the two nucleons, the placement
of the $\hQ_i$ immediately allows one to find the momenta along each nucleon
line. Equation (\ref{eq:divdmu}) is an immediate consequence of the
Ward-Takahashi identity for the nucleon current operator
$\Gamma_i^\mu$~\cite{Takahashi,WeinbergI}, i.e.,
\begin{equation}
  k_\mu \Gamma^\mu_i =  \hQ_i t^{-1}_i-t^{-1}_i \hQ_i~,
  \label{eq:singleWTI}
\end{equation}
applied to Eq.~(\ref{eq:dgauge}). Explicitly, with $G_0$ specified as in
(\ref{eq:G0conv}), we have
\begin{widetext}
\begin{align}
\hQ G_0^{-1}-G_0^{-1} \hQ
&=  [\hQ_1 t_1^{-1}-t^{-1}_1 \hQ_1]\circ t^{-1}_2
     +  t_1^{-1}\circ [\hQ_2 t_2^{-1}-t^{-1}_2 \hQ_2]
\nonumber\\
&= [Q_1 t_1^{-1}(p_1)-t^{-1}_1(p_1-k) Q_1]\circ t^{-1}_2(p_2)
  + t_1^{-1}(p_1)\circ [Q_2 t_2^{-1}(p_2)-t^{-1}_2(p_2-k) Q_2]
\end{align}
for a two-nucleon system where $p_1$ and $p_2$ are the initial four-momenta of
nucleons 1 and 2, respectively, i.e., the first term results from the photon
being emitted by nucleon 1 and for the second term, it is being emitted by
nucleon 2.

In the same schematic notation, the four-divergence of the interaction current
$V^\mu$ then is simply~\cite{H97,GrossRiska}
\begin{align}
k_\mu V^\mu &= V\hQ-\hQ V
\nonumber\\
&= V(p'_1,p'_2;p_1-k,p_2)Q_1+V(p'_1,p'_2;p_1,p_2-k)Q_2-Q_1V(p'_1+k,p'_2;p_1,p_2)-Q_2V(p'_1,p'_2+k;p_1,p_2)
~,
\label{eq:divVmu}
\end{align}
where the arguments of $V$ are nucleon momenta and the momentum dependence of
the interaction current is
\begin{equation}
  V^\mu=V^\mu(k,p_1',p_2';p_1,p_2)~,
  \qquad\text{with}\qquad p'_1+p'_2+k=p_1+p_2~,
  \label{eq:VmuMomenta}
\end{equation}
i.e., the momenta $p_1,p_2$ and $p_1',p_2'$ are those of the incoming and
outgoing nucleons, respectively. Whether the charge operators $Q_i$ in
(\ref{eq:divVmu}) pertain to incoming or outgoing nucleons is clear from where
the $Q_i$ are placed in the equation. In other words, if placed on the right of
$V$, $Q_i$ describes the charge of the incoming nucleon $i$, and if placed on
the left, it describes the charge of the outgoing nucleon $i$. Placing the
charge operators in this manner is necessary since they interact with the
isospin dependence of the interaction $V$.

The four-divergence of $J^\mu$ then follows as
\begin{align}
k_\mu J^\mu
&=\left(\hQ G_0^{-1}-G_0^{-1} \hQ\right)G_0 V+VG0\left(\hQ G_0^{-1}-G_0^{-1} \hQ\right)+V\hQ-\hQ V
-VG_0 (\hQ G_0^{-1}-G_0^{-1} \hQ) G_0 V
\nonumber\\
&=-G_0^{-1}\hQ G_0 V+VG_0\hQ G_0^{-1}-VG_0 \hQ V+V \hQ G_0 V~.
\label{eq:4divJmu}
\end{align}
For the entire current, we then find
\begin{equation}
k_\mu B^\mu = (1+TG_0)\left(VG_0\hQ G_0^{-1}-G_0^{-1}\hQ G_0 V-VG_0 \hQ V+V \hQ G_0 V\right)(G_0T+1)~,
\label{eq:deriveWTI}
\end{equation}
and thus finally, using (\ref{eq:NNLS}),
\begin{equation}
k_\mu B^\mu=TG_0\hQ G_0^{-1}-G_0^{-1}\hQ G_0 T~.
\label{eq:WTI}
\end{equation}
This is the correct generalized Ward-Takahashi identity~\cite{Kazes} for the
bremsstrahlung current providing a conserved current for external on-shell
nucleons. In a more explicit notation, using the same arguments for $T$ and
$B^\mu$ as for $V$ and $V^\mu$, respectively, in Eqs.~(\ref{eq:divVmu}) and
(\ref{eq:VmuMomenta}), this reads
\begin{align}
k_\mu B^\mu &=
T(p'_1,p'_2;p_1-k,p_2)\,t_1(p_1-k)\,Q_1\,t_1^{-1}(p_1)
  + T(p'_1,p'_2;p_1,p_2-k)\,t_2(p_2-k)\,Q_2\,t_2^{-1}(p_2)
\nonumber\\[1ex]
&\mbox{}\qquad
  -t_1^{-1}(p'_1)\,Q_1\,t_1(p'_1+k)\, T(p'_1+k,p'_2;p_1,p_2)
  -t_2^{-1}(p'_2)\,Q_2\,t_2(p'_2+k)\, T(p'_1,p'_2+k;p_1,p_2)~.
  \label{eq:WTBmu}
\end{align}
\end{widetext}
The inverse nucleon propagators $t_i^{-1}(p)$ appearing here ensure that this
four-divergence vanishes (i.e., that the bremsstrahlung current is conserved)
if all external nucleon legs are on-shell.

\subsection{Covariant three-dimensional reduction}\label{sec:3dimred}

To calculate any reaction amplitude in a full four-dimensional framework is a
daunting numerical task. In practical applications of relativistic reaction
theories, therefore, one often employs three-dimensional reductions that
eliminate the energy variable from loop integrations in a covariant manner
leaving only integrations over the components of three-momenta. Many such
reduction schemes can be found in the literature
\cite{BlankenbeclerSugar,E74,CJ89,GrossText}. Our results presented below hold
true for any reduction scheme that puts both nucleons in loops on their
respective energy shells.

For hadronic reactions, the primary technical constraint to be satisfied by any
three-dimensional reduction is the preservation of covariance and
(relativistic) unitarity. For reactions involving electromagnetic interactions,
there is the additional constraint of gauge invariance. This is a non-trivial
constraint since the reduction scheme, in general, will destroy gauge
invariance as a matter of course. Hence, to restore it, one must introduce
additional current mechanisms as part of the reduction prescription for
photoprocesses. As we shall see, this cannot be done in a unique manner because
gauge invariance does not constrain transverse current contributions.

For the $NN$ problem, three-dimensional reductions result from replacing the
free two-nucleon propagator $G_0$ by one containing a $\delta$ function that
eliminates the energy integration in loops. In the following, we make the
replacement
\begin{equation}
  G_0 \to g_0
  \label{eq:G0red}
\end{equation}
to indicate that the internal integration is a three-dimensional one over the
three-momentum of the loop. To obtain on-the-energy-shell \emph{integral
equations} from the Bethe-Salpeter equations (\ref{eq:NNLS}) when using this
reduction, the external nucleon legs must be taken on shell as well. This
provides the reduced integral equations
\begin{equation}
  t=v+vg_0t =v+tg_0 v~,
  \label{eq:BSred}
\end{equation}
where lower-case letters $v$ and $t$ (instead of $V$ and $T$, respectively)
signify that all nucleons --- internal and external --- are on their energy
shell. However, when considering below the gauge invariance of the
bremsstrahlung current that results from the reduction (\ref{eq:G0red}), we
require fully off-shell $T$ matrices. They can be obtained from iterated
versions of (\ref{eq:NNLS}) which is then subjected to the reduction
(\ref{eq:G0red}), producing
\begin{equation}
  T = V + V(g_0+g_0tg_0)V~,
  \label{eq:Toffshell}
\end{equation}
where all external nucleons may be considered off-shell. This off-shell $T$,
thus, is obtained by quadratures from the integral-equation on-shell solution
$t$. However, this $T$ is \emph{not} the same as the solution of the full
four-dimensional Bethe-Salpeter scattering equation (\ref{eq:NNLS}) even though
we use the same notation to keep matters simple, even at the risk of inviting
confusion. The $T$'s appearing in the following context \emph{always} refer to
fully off-shell or half off-shell versions of the $T$ matrix defined by the
quadrature formula (\ref{eq:Toffshell}).

The previous proof of gauge invariance of the full four-dimensional formalism
given in Sec.~\ref{sec:4dimgauge} shows that gauge invariance depends on an
intricate interplay of all hadronic reaction mechanisms. Any approximation,
therefore, will in general destroy gauge invariance. Hence, we expect that
simply subjecting the full bremsstrahlung current $B^\mu$ of
Eq.~(\ref{eq:BmuGeneric}) to the reduction prescription (\ref{eq:G0red}) will
\emph{not} retain gauge invariance, and that, therefore, additional steps will
be necessary to ensure gauge invariance. Since, from a formal point of view,
\emph{any} modification of a given current can, without lack of generality,
always be expressed by adding an extra current, we may write the
\emph{gauge-invariant} current $B^\mu_r$ that results from a judicious
adaptation of the three-dimensional reduction procedure to $B^\mu$ of
Eq.~(\ref{eq:BmuGeneric}) in the form
\begin{equation}
  B^\mu \to B^\mu_r =\Big[(TG_0+1)J^\mu(1+G_0T)\Big]_{\text{red}} + X^\mu_\GIP~.
  \label{eq:redBmuAnsatz}
\end{equation}
The first term on the right-hand side, $[\cdots]_{\text{red}}$, schematically
denotes the necessary modifications of the hadronic mechanisms in $B^\mu$
itself and the last term, $X^\mu_\GIP$, is the additional gauge-invariance
preserving (GIP) current that is to be determined to make $B^\mu_r$ gauge
invariant. In other words, we \emph{demand} that
\begin{equation}
  k_\mu B^\mu_r =TG_0\hQ G_0^{-1}-G_0^{-1}\hQ G_0 T~,
  \label{eq:DivBmuRed}
\end{equation}
i.e., that the four-divergence of the reduced current $B^\mu_r$ remains
identical in form to the generalized WTI of Eq.~(\ref{eq:WTI}) and we are going
to ensure this by choosing $X^\mu_\GIP$ accordingly after having determined
$[\cdots]_{\text{red}}$ in (\ref{eq:redBmuAnsatz}). [We repeat here  that ---
 completely consistent with the covariant three-dimensional reduction --- the
 off-shell $T$'s of (\ref{eq:DivBmuRed}), and of the subsequent equations in
 this section, are those defined by the off-shell extension
 (\ref{eq:Toffshell}) of (\ref{eq:BSred}), and \emph{not} the solutions of the
 original Bethe-Salpeter equations (\ref{eq:NNLS}) that appear in
 (\ref{eq:WTI}); see also the discussion surrounding Eq.~(\ref{eq:Toffshell}).]

%
\begin{figure}[t!]\centering
  \includegraphics[width=.8\columnwidth,clip=]{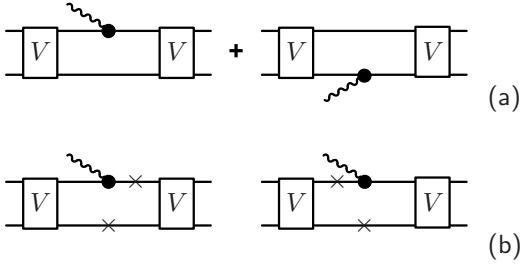}
  \caption{\label{fig:VdV}%
  (a) Sum of box graphs describing Eq.~(\ref{eq:VdVbox}), with rectangles
  labeled $V$ subsuming all mechanisms of the $NN$ interaction. All internal
  nucleon lines are off-shell, in general. ~(b) The two possibilities of
  putting nucleons on-shell in the internal loop of the first diagram of (a),
  as indicated by ``$\times$'' on the corresponding nucleon line, leaving one
  nucleon at the photon vertex off shell. Analogous diagrams can be drawn for
  the corresponding second diagram.}
\end{figure}
%

As a first step, we employ the reduction for the external $G_0$ factors and
write
\begin{equation}
  B^\mu_r = (Tg_0+1)\tJ^\mu_r (1+g_0T)+X^\mu_\GIP%
  ~,
  \label{eq:BmuRed0}
\end{equation}
where $\tJ_r^\mu$ is the reduced form of the basic production current $J^\mu$
of (\ref{eq:JmuModified}). For its determination, we note that one cannot
simply employ the reduction (\ref{eq:G0red}) for every $G_0$ appearing in
(\ref{eq:JmuModified}) since this produces unphysical mechanisms. The box-graph
contribution
\begin{equation}
  b^\mu = VG_0d^\mu G_0 V
  \label{eq:VdVbox}
\end{equation}
shown in Fig.~\ref{fig:VdV}(a), that appears as a subtraction in
(\ref{eq:DmuDefined}), \emph{cannot} be reduced in the form
\begin{equation}
VG_0d^\mu G_0 V \to Vg_0d^\mu g_0 V\qquad\mbox{(unphysical)}
\label{eq:VdVunphys}
\end{equation}
since this would have the bremsstrahlung photon emerging from intermediate
on-shell nucleons which is not possible for a physical photon.  At least one of
the nucleon legs at the photon vertex --- either the incoming or the outgoing
one --- must remain off-shell.  The two possibilities of doing that allowing
the production of physical photons are
\begin{equation}
VG_0d^\mu G_0 V \to \begin{cases}
VG_0d^\mu g_0 V~,\\[1ex]
Vg_0d^\mu G_0 V~,
\end{cases}
\end{equation}
as shown in Fig.~\ref{fig:VdV}(b). Since there is nothing that suggests that
one choice is to be preferred over the other, we allow for both and thus make
the replacement
\begin{equation}
b^\mu \to b^\mu_r= \lambda_i VG_0d^\mu g_0 V+\lambda_f Vg_0d^\mu G_0 V~,
\end{equation}
where the constant factors are constrained by
\begin{equation}
  \lambda_i+\lambda_f =1
  \label{eq:lambdasum}
\end{equation}
to avoid double counting. The symmetric choice would be
$\lambda_i=\lambda_f=1/2$, of course, but we want to allow here more
flexibility for reasons given below.

%
\begin{figure*}[t!]\centering
\includegraphics[width=.8\textwidth,clip=]{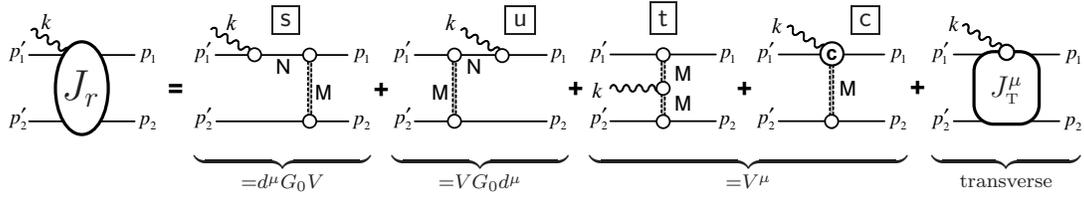}
\caption{\label{fig:Jfigred}%
Reduced basic bremsstrahlung production current $J^\mu_r$ of
Eq.~(\ref{eq:modJmu}) resulting from the covariant three-dimensional reduction
(see Sec.~\ref{sec:3dimred}) for nucleon-nucleon interactions based on
single-meson exchanges. Time proceeds from right to left. The mesons $M$ taken
into account in the present work and in Ref.~\cite{NH09} are $\pi$, $\eta$,
$\rho$, $\omega$, $\sigma$, and $a_0$ (formerly $\delta$). The nucleonic
current corresponds to the first two diagrams on the right-hand side and the
meson-exchange current is depicted by the third diagram. The first four
diagrams respectively labeled s, u, t, and c correspond to the \emph{complete}
gauge-invariant description for the process $NM\to N\gamma$ for the upper
nucleon line, with the labels s, u, and t alluding to the kinematic situations
described by the corresponding Mandelstam variables (see Fig.~\ref{fig:gammapi}
and the corresponding discussion in Sec.~\ref{subsec:singlemeson}). The fourth
diagram contains the $N M\to N \gamma$ four-point contact current $M^\mu_\IC$
discussed in Sec.~\ref{sec:phenIntCurrent}, labeled ``c" in the diagram. The
correct gauge-invariance treatment of this diagram was found to be crucial in
reproducing the KVI data in Ref.~\cite{NH09}. The diagrams corresponding to s,
u, and c for the lower nucleon line are suppressed. The last diagram (labeled
$\JT^\mu$) subsumes the transverse five-point currents of (\ref{eq:bmuT1}).
Transverse transition-current contributions are also subsumed in this diagram;
examples are the $\gamma\rho\pi$, $\gamma\omega\pi$, and $\gamma N\Delta$
transition currents depicted in Figs.~\ref{fig:JMeson} and
\ref{fig:DeltaCurrents}, respectively. Antisymmetrization of identical nucleons
is implied.}
\end{figure*}
%

In detail, the reduced bremsstrahlung current thus reads
\begin{widetext}
\begin{equation}
  B_r^\mu =(Tg_0+1)\big[d^\mu G_0 V + VG_0 d^\mu +V^\mu
  +D^\mu_r\big] (1+g_0T) +X^\mu_\GIP~,
\end{equation}
where
\begin{equation}
  D^\mu_r = T^\mu_r -b^\mu_r
\end{equation}
is the reduced form of (\ref{eq:DmuDefined}), with $T^\mu_r$ denoting the
reduced form of the transverse current $T^\mu$ satisfying $k_\mu T^\mu_r=0$.
Note that the $G_0$'s appearing in the $d^\mu$ terms cannot be reduced to
$g_0$'s for the same reason that the reduction (\ref{eq:VdVunphys}) is not
possible. Evaluating the four-divergence of this expression provides
\begin{align}
  k_\mu B^\mu_r &= (Tg_0+1)\left[\left(\hQ G_0^{-1}-G_0^{-1}\hQ\right) G_0 V
  + VG_0 \left(\hQ G_0^{-1}-G_0^{-1}\hQ\right) +V\hQ-\hQ V
  - k_\mu b^\mu_r  \Big)
 \right] (1+g_0T) +k_\mu X^\mu_\GIP
 \nonumber\displaybreak[0]\\
 &=-(Tg_0+1) G_0^{-1} \hQ G_0 T +T G_0 \hQ G_0^{-1} (1+g_0T)
 -(Tg_0+1)k_\mu b^\mu_r   (1+g_0T)  +k_\mu X^\mu_\GIP
 \nonumber\displaybreak[0]\\
 &=T G_0 \hQ G_0^{-1} - G_0^{-1} \hQ G_0 T
 -(Tg_0+1)k_\mu b^\mu_r   (1+g_0T) +k_\mu X^\mu_\GIP~.
 \label{eq:BmuRedDiv0}
\end{align}
\end{widetext}
It was used here that
\begin{equation}
  T G_0 \hQ G_0^{-1}g_0 T=0
  \quad\text{and}\quad
  T g_0 G_0^{-1} \hQ G_0 T=0
  \end{equation}
vanish identically. For the proof, consider
\begin{equation}
  g_0 G_0^{-1} \hQ G_0 \to \frac{\Lambda_1\Lambda_2}{s-(\varepsilon_1+\varepsilon_2)^2}\circ \left[t_1^{-1}(p)Q_1 t_1(p+k)\right]~,
\end{equation}
where the right-hand side here provides one generic contribution (stripped of
all extraneous factors) contained in the expression on the left. The ``$\circ$"
symbol indicates the remaining three-momentum loop integration. The variable
$s$ is the squared total energy of the system and the $\varepsilon_i$ are the
individual on-shell energies of the two on-shell nucleons in the loop
[indicated by the symbol ``$\times$'' in Fig.~\ref{fig:VdV}(b)]. The
$\Lambda_i$ are the positive-energy projectors of the nucleons. The energy
component of $p=(\varepsilon_1,\mathbf{p})$ is on-shell and thus
\begin{equation}
 \Lambda_1(\mathbf{p})\, t_1^{-1}(p)= \frac{p^2-m^2}{2m} \Lambda_1(\mathbf{p}) =0~.
\end{equation}
Because of $p^2=m^2$, this term \emph{always} vanishes for any $\mathbf{p}$.
Therefore, even if $s-(\varepsilon_1+\varepsilon_2)^2$ should vanish as well
for some $\mathbf{p}$, this cannot be compensated, i.e., the limit of the
corresponding $\frac{0}{0}$ situation is zero. This proves that $Tg_0
G_0^{-1}\hQ G_0 T=0$; the proof for $ T G_0\hQ G_0^{-1}g_0T=0$ follows in a
similar fashion.

The first two terms in the last line of (\ref{eq:BmuRedDiv0}) already provide
the complete four-divergence (\ref{eq:DivBmuRed}) necessary for the
gauge-invariance condition to hold true. It follows then that the last two
terms must vanish,
\begin{equation}
k_\mu X^\mu_\GIP -  (Tg_0+1)k_\mu b^\mu_r   (1+g_0T)\stackrel{!}{=}0~.
\end{equation}
The four-divergence of $X^\mu_\GIP$ thus is constrained by
\begin{equation}
k_\mu X^\mu_\GIP = (Tg_0+1)k_\mu b^\mu_{\textsc{l}}   (1+g_0T)~,
\label{eq:XmuConstr}
\end{equation}
where $b^\mu_{\textsc{l}}$ describes the longitudinal pieces of the reduced
box-graph current $b^\mu_r$. Hence, without lack of generality, we may write
\begin{equation}
  X^\mu_\GIP + (Tg_0+1) D^\mu_r   (1+g_0T) = (Tg_0+1) \JT^\mu   (1+g_0T)
\end{equation}
where $\JT^\mu$ is the purely \emph{transverse} current
\begin{equation}
\JT^\mu= T^\mu_r-\lambda_i VG_0d_{\textsc{t}}^\mu g_0 V-\lambda_f Vg_0d_{\textsc{t}}^\mu G_0 V ~;
\label{eq:bmuT1}
\end{equation}
$d_{\textsc{t}}^\mu$ here only contains the transverse pieces of the nucleon
currents as they appear in $d^\mu$. As far as gauge invariance of $B^\mu_r$ is
concerned, the current $J^\mu_{\textsc{t}}$ is irrelevant. Therefore, the
parameters $\lambda_i$ and $\lambda_f$ may now also be treated as independent
parameters, unconstrained by (\ref{eq:lambdasum}).

To summarize the present results obtained for the three-dimensional reduction,
in this approximation the gauge-invariant bremsstrahlung current reads
\begin{equation}
  B^\mu_r = (Tg_0+1) J^\mu_r (1+g_0T)~,
  \label{eq:modBmu}
\end{equation}
where the reduced basic production current is given as
\begin{equation}
  J^\mu_r =d^\mu G_0 V + VG_0 d^\mu +V^\mu + J^\mu_{\textsc{t}}~.
  \label{eq:modJmu}
\end{equation}
How one chooses $J^\mu_{\textsc{t}}$ in an application is not fixed by the
formalism, beyond the generic form given in (\ref{eq:bmuT1}).  Note that the
generic graphical structure depicted in Fig.~\ref{fig:Jgeneric} remains valid
also for $J^\mu_r$, with the last graph labeled $D^\mu$ on the right-hand side
of the figure depicting now the transverse five-point current $\JT^\mu$. For
the specific case of  $NN$ interactions based on single-meson exchanges only,
the reduced basic production current $J^\mu_r$ of (\ref{eq:modJmu}) is
illustrated diagrammatically in more detail in Fig.~\ref{fig:Jfigred}.

Let us add some remarks here. Even though the procedure to preserve gauge
invariance was presented here in terms of an additional \emph{ad hoc} current
$X^\mu_\GIP$, the derivation shows that, rather than adding a current, the
application of the three-dimensional reduction procedure to the current $B^\mu$
really amounts to \emph{dropping} (at least part of) the reduced contribution
from $b^\mu$ that was necessary in the full four-dimensional treatment to
prevent double counting of the $T G_0 d^\mu G_0 T$ contribution. The particular
form of $\JT^\mu$ of (\ref{eq:bmuT1}) follows from exploiting the constraint
(\ref{eq:XmuConstr}) to the extent to which it is possible since the gauge
invariance cannot constrain transverse contributions. Our applications reported
here, in Sec.~\ref{sec:Application}, and in \cite{NH09} suggest that the choice
$\lambda_i=\lambda_f=0$ yields by far the best numerical results. in view of
the fact that the double-counted term $T G_0 d^\mu G_0 T$ in the full
four-dimensional formulation of Sec.~\ref{sec:4dimformalism} now appears in
(\ref{eq:modBmu}) as two distinctly different contributions $T G_0 d^\mu g_0 T
+T g_0 d^\mu G_0 T$, there are now no longer any terms counted doubly and thus
the corresponding subtraction is no longer necessary either. The finding that
$\lambda_i=\lambda_f=0$ yields the best numerical results is completely
consistent with this fact. In our calculations, therefore, $\JT^\mu$ only
contains transverse contributions from electromagnetic $\gamma\rho\pi$,
$\gamma\omega\pi$, and $\gamma N \Delta$ transitions (for the latter, see
discussion in Sec.~\ref{sec:delta}).

\subsection{\boldmath Local gauge invariance: Constraining subprocesses}\label{subsec:singlemeson}

The gauge-invariance constraints in terms of the generalized Ward-Takahashi
identities, either in the form (\ref{eq:WTI}) for the full amplitude or as
(\ref{eq:DivBmuRed}) for the reduced one, are just formal constraints, of
course, since the only physically relevant --- i.e., measurable
--- ramification is current conservation when all external nucleons of the
bremsstrahlung process are on their respective energy shells. It is for this
reason that the current-conservation constraint,
\begin{equation}
  k_\mu B^\mu =0 \qquad \text{(external nucleons on-shell)}~,
\end{equation}
called \emph{global} gauge invariance, is the only constraint that is
implemented in many reaction models of photoprocesses. We would like to
advocate, however, that using this as the sole constraint is not enough if one
aims at providing a consistent \emph{microscopic} description of the
photoreaction at hand. It was shown in the preceding sections that the gauge
invariance of the total bremsstrahlung amplitude hinges in an essential way on
each subprocess providing its correct current share to ensure the gauge
invariance of the entire current --- and thus ultimately provide a conserved
current. This is only possible if the current associated with each subprocess
satisfies its own generalized Ward-Takahashi identity, as exemplified here by
Eq.~(\ref{eq:divdmu}) for the propagators and by Eq.~(\ref{eq:divVmu}) for the
interaction current. Thus, imposing \emph{local} gauge invariance, i.e.,
imposing consistent \emph{off-shell} constraints of this kind for all
subprocesses in a microscopic description of the reaction at hand will then
automatically ensure that the total amplitude will satisfy a generalized WTI of
its own. This not only means that it will indeed satisfy the physical
constraint of a conserved current, but beyond that it will ensure that if the
process at hand will be used as a subprocess for another, larger reaction, it
will \emph{automatically} provide the correct contribution to make the larger
process gauge invariant as well.

\begin{figure}[t!]\centering
  \includegraphics[width=.9\columnwidth,clip=]{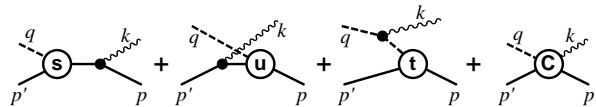}
  \caption{\label{fig:gammapi}%
  Generic pion photoproduction diagrams $\gamma +N \to \pi +N$. Time proceeds
  from right to left. If read from left to right, this corresponds to the
  pion-capture reaction as it appears in the first four diagrams on the
  right-hand side of Fig.~\ref{fig:Jfigred} along the upper nucleon line. The last
  diagram here labeled ``c'' corresponds to the interaction current arising
  from attaching the photon to the interior of the $\pi NN$ vertex. The
  properties of this contact-type four-point current $M^\mu_\IC$ are essential to
  render the entire amplitude gauge invariant, with the necessary constraint
  equation given by (\ref{eq:IntCurrentGaugeCond}). At the tree-level for
  undressed hadrons, this term reduces to the Kroll-Ruderman term.}
\end{figure}

In this respect, we draw attention to the fact that the reaction dynamics
depicted by the first four diagrams on the right-hand side of
Fig.~\ref{fig:Jfigred} correspond to the capture reaction $N M \to N \gamma$
where the meson $M$ is emitted by the spectator nucleon depicted by the lower
nucleon line. The corresponding time-reversed equivalent meson-production
process is shown in Fig.~\ref{fig:gammapi} for the example of pion production.
The current amplitude for this process can always be broken down into four
generic contributions (see, for example, \cite{H97}),
\begin{equation}
  M^\mu = M^\mu_s +M^\mu_u+M^\mu_t+M^\mu_\IC~,
\end{equation}
where the indices $s$, $u$, and $t$ allude to the Mandelstam variables
describing the kinematic situations of the corresponding diagrams in the
figure. The corresponding interaction current, in particular, is obtained by
attaching the photon to the inner workings of the meson-nucleon-nucleon vertex
$F$ according to
\begin{equation}
M^\mu_\IC = -\left\{F\right\}^\mu~.
\end{equation}
If we demand now local gauge invariance, the current $M^\mu$ must satisfy an
(off-shell) generalized WTI. In addition to the trivial contributions resulting
from the propagator WTIs for the currents associated with the external legs of
$M^\mu_s$, $M^\mu_u$, and $M^\mu_t$, similar to (\ref{eq:singleWTI}), this
means in particular, that the interaction current must satisfy~\cite{H97}
\begin{equation}
  k_\mu M^\mu_\IC =   Q_N F_u +Q_M F_t - F_s Q_N~,
  \label{eq:IntCurrentGaugeCond}
\end{equation}
where $F_x$ denotes the meson-nucleon-nucleon vertices $F$, with the subscripts
$x=s,u,t$ corresponding to Mandelstam variables of the respective kinematical
situation of the vertices in \mbox{$s$-,} \mbox{$u$-,} and $t$-channel
contributions, as depicted in Fig.~\ref{fig:gammapi}. $Q_M$ and $Q_N$ are the
charge operators of the meson and the nucleon, respectively. Since the charge
operators interact with the isospin dependence of the vertex $F$, their
placement before or after $F_x$ is significant to ensure charge conservation.
Note that Eq.~(\ref{eq:IntCurrentGaugeCond}) is the exact analog of
Eq.~(\ref{eq:divVmu}) for the present application, with Eq.~(\ref{eq:divVmu})
providing the four-divergence for a five-point interaction current for an
outgoing photon and Eq.~(\ref{eq:IntCurrentGaugeCond}) constraining the
four-point interaction current for an incoming photon. Using the $\hQ$ notation
introduced in Eq.~(\ref{eq:Qhatintro}), this may be made more obvious by
writing Eq.~(\ref{eq:IntCurrentGaugeCond}) as
\begin{equation}
  k_\mu M^\mu_\IC = \left(\hQ_N +\hQ_M\right)F -F \hQ_N~,
  \label{eq:IntCurrConstraint}
\end{equation}
where the charge operators of the final hadrons appear on the left of $F$ and
that of the initial hadron on its right, as it is appropriate for the
interaction current of any meson-production process $\gamma N\to NM$ similar to
Fig.~\ref{fig:gammapi}; for the reverse process $NM\to N \gamma$, one needs to
change $k\to-k$ and exploit the isospin dependence of the vertex to write
$\hQ_MF=-F\hQ_M$ since the meson changes from being outgoing to incoming.

%
%

%
\begin{figure*}[t!]\centering
  \includegraphics[width=.6\textwidth,clip=]{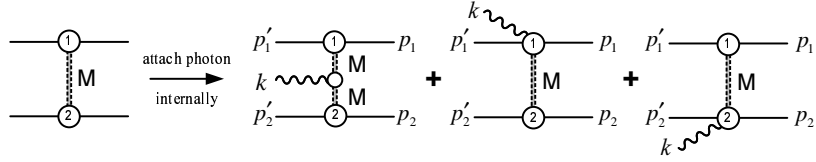}
  \caption{\label{fig:Vmudetail}%
  Generic structure of interaction current $V^\mu_\MEC$, Eq.~(\ref{eq:Vsmu}), for the
  single-meson exchange contribution $V_\MEC$ of the $NN$ interaction depicted on
  the left. Time proceeds from right to left. The three diagrams on the
  right-hand side comprise the meson-exchange current and the two
  interaction-current contributions $M^\mu_i$, $i=1,2$, where the photon
  interacts with the interiors of the respective vertices for nucleons 1 and
  2.}
\end{figure*}
%

Demanding consistency of the microscopic dynamics across various reactions
means that the entire $N M \to N \gamma$ subprocess in Fig.~\ref{fig:Jfigred}
must satisfy the \emph{same} generalized Ward-Takahashi identity as the
amplitude $M^\mu$ itself, except for trivial modifications arising from the
fact that this is the time-reversed process. This must be true for any one of
the exchanged mesons --- whether scalar, pseudoscalar, or vector
---  in the bremsstrahlung process, not just for the pion example shown in
Fig.~\ref{fig:gammapi}.

To illustrate in more detail how the requirement of local gauge invariance ties
together the various current mechanisms, let us consider a
single-meson-exchange contribution $V_\MEC$ to the full $NN$ potential $V$.
Generically, we may write
\begin{equation}
  V_\MEC = F_1 t_M F_2~,
\end{equation}
where the $F_i$ are the meson-nucleon-nucleon vertices for nucleon $i=1,2$
(including all coupling operators and isospin dependencies) and $t_M$ describes
the propagator for the exchanged meson $M$. Graphically, the process is given
in the diagram on the left-hand side of Fig.~\ref{fig:Vmudetail}. If we now
attach an outgoing photon to all internal mechanism of $V_\MEC$, we obtain the
three diagrams on the right-hand side of the figure that comprise the
corresponding $NN$ interaction current, $V_\MEC^\mu
=-\left\{V_\MEC\right\}^\mu$. This current may be written as
\begin{align}
  V_\MEC^\mu  &= -F_1 \left\{t_M\right\}^\mu F_2 -\left\{F_1\right\}^\mu t_M F_2 -F_1t_M \left\{F_2\right\}^\mu
  \nonumber\\
  &= F_1 t_M\, \Gamma^\mu_M\, t_M F_2 +M^\mu_1 t_M F_2 +F_1 t_M M^\mu_2~,
  \label{eq:Vsmu}
\end{align}
where $M_i^\mu  = -\left\{F_i\right\}^\mu$ is the four-point interaction
current for the vertex $i=1,2$ and
\begin{equation}
  -\left\{t_M\right\}^\mu = t_M \, \Gamma^\mu_M\, t_M
\end{equation}
produces the current operator $\Gamma^\mu_M$ for the exchanged meson that
satisfies the  single-particle WTI~\cite{Takahashi,WeinbergI}
\begin{equation}
  k_\mu \Gamma^\mu_M = \hQ_M t^{-1}_M-t^{-1}_M \hQ_M~.
\end{equation}
Now, assuming without lack of generality that four-momentum and charge
flow from vertex 2 to vertex 1 in Fig.~\ref{fig:Vmudetail}, the analogs of
(\ref{eq:IntCurrConstraint}) read
\begin{subequations}\label{eq:intMmuAll}
\begin{align}
  k_\mu M_1^\mu &=F_1\left(\hQ_1+\hQ_M \right) -\hQ_1 F_1~,
  \label{eq:intMmu1}
  \\
  k_\mu M_2^\mu &= F_2 \hQ_2- \left(\hQ_M+\hQ_2\right)F_2~,
  \label{eq:intMmu2}
\end{align}
\end{subequations}
and thus
\begin{align}
  k_\mu V^\mu_\MEC &=F_1 t_M\, \hQ_M F_2-F_1 \hQ_M t_M F_2
  \nonumber\\
  &\quad\mbox{}
  +F_1\left(\hQ_1+\hQ_M \right) t_M F_2-\hQ_1 F_1 t_M F_2
  \nonumber\\
  &\quad\mbox{}
  +F_1 t_M F_2 \hQ_2-F_1 t_M \left(\hQ_M+\hQ_2\right)F_2
  \nonumber\\
  &=F_1 t_M F_2 \hQ_1-\hQ_1 F_1 t_M F_2
  \nonumber\\
  &\quad\mbox{}
  +F_1 t_M F_2 \hQ_2-\hQ_2 F_1 t_M F_2
  \nonumber\\
  &= V_\MEC \hQ -\hQ V_\MEC
\end{align}
which is precisely the gauge-invariance constraint (\ref{eq:divVmu}) for this
particular contribution to $V$. In this consistent microscopic treatment of all
subprocesses, therefore, the constraints \emph{local} gauge invariance places
on the four-point interaction currents $M^\mu_\IC$ of meson-production
processes translate seamlessly into the corresponding constraints on five-point
currents of the bremsstrahlung process. The essential step here is to ensure
the validity of (\ref{eq:intMmu1}) for the four-point current in the diagram
labeled ``c'' in Fig.~\ref{fig:Jfigred} and of (\ref{eq:intMmu2}) for its
counterpart for the lower nucleon line (not shown in Fig.~\ref{fig:Jfigred}).
These relations must be true for each of the exchanged mesons.

The problem of how to ensure the validity of the constraints
(\ref{eq:IntCurrentGaugeCond}) or (\ref{eq:intMmuAll}) has been studied
extensively by the present authors and their collaborators for the equivalent
photoproduction processes, and based on the original ideas presented in
Refs.~\cite{H97,HBMF98}, a general prescription was given in Ref.~\cite{HNK06}
that is applicable just as well for the most general case of explicit
final-state interactions with completely dressed hadrons as it is for
phenomenological vertex functions. The necessary extension in the context of
bremsstrahlung --- to account for the virtual nature of the incoming and
outgoing nucleons and the exchanged meson at the four-point vertex --- is
accomplished following the work of Ref.~\cite{NOH06}.

For our present applications, and those of Ref.~\cite{NH09}, based on
single-meson exchanges containing phenomenological meson-nucleon-nucleon form
factors, the details of the four-point currents $M^\mu_1$ and $M^\mu_2$ and of
the complete descriptions of the corresponding diagrams depicted in
Fig.~\ref{fig:Vmudetail} were already given in Ref.~\cite{NH09}; we will not
repeat them here. We emphasize, however, that the inclusion of this
interaction-type current, with the correct gauge-invariance dynamics that
ensure the validity of Eqs.~(\ref{eq:intMmuAll}), is essential to bringing
about the good quality of our bremsstrahlung results, in particular, the good
description of the high-precision KVI data~\cite{KVI02,KVI04} reported in
Ref.~\cite{NH09}. This is a novel feature of our approach that resolves a
longstanding discrepancy between the data and their theoretical description.

\subsection{\boldmath Interaction current $V^\mu$ for phenomenological $NN$ interactions:
A systematic gauge-invariance-\ preserving
approximation}\label{sec:phenIntCurrent}

The previous discussions show that the gauge-invariance conditions
(\ref{eq:divVmu}) and (\ref{eq:intMmuAll}) for the respective interaction
currents are of crucial importance to ensure an overall bremsstrahlung
amplitude that satisfies gauge invariance. The understanding here is that the
detailed reaction mechanisms that go into providing the details of these
interaction currents are completely known and that therefore it would be
straightforward to make sure that the corresponding gauge-invariance conditions
are indeed satisfied. This is generally the case only for $NN$ interactions
based on single-meson exchanges between nucleons \emph{without} any
phenomenological form factors. And if one employs phenomenological form
factors, one must resort to the prescriptions given in
Refs.~\cite{H97,HBMF98,HNK06,NOH06} to ensure Eqs.~(\ref{eq:intMmuAll}), as
discussed in the preceding section. However, beyond problems associated with
phenomenological form factors, it is conceivable that some details would not be
available in some instances either because the microscopic coupling of the
photon to the internal dynamics of the interaction would not be feasible or
sensible or because it would be too complicated for practical applications.
Examples of the former would be $NN$ interactions based on position-space
methods where there are no (momentum-dependent) exchange mechanisms that permit
the coupling of a photon in a dynamically meaningful way. Examples of the
latter might be baryon contributions beyond the nucleon since they require
box-type $NN$ contributions with intermediate non-nucleonic baryonic
contributions that might be too cumbersome to be treated with explicit photon
couplings (see Sec.~\ref{sec:delta} for $N\Delta$ and/or $\Delta\Delta$ systems
and the resulting graphical structures depicted in Fig.~\ref{fig:DeltaBoxes}).

Nevertheless, each of the corresponding interaction-current contributions
\emph{must} satisfy its appropriate gauge-invariance condition otherwise the
gauge invariance of the entire amplitude breaks down. We will show here how one
can ensure that gauge invariance is preserved for any phenomenological
contribution to the $NN$ interaction by constructing a phenomenological
interaction current that satisfies appropriate constraints.

To this end let us assume the total $NN$ interaction $V$ can be split up into
$n$ \emph{independent} contributions,
\begin{equation}
  V = V_1 +V_2+\cdots +V_n~.
\end{equation}
Formally coupling a photon to each contribution according to  $V_i^\mu
=-\{V_i\}$, the total interaction current $V^\mu$ then breaks down accordingly
into $n$ independent contributions,
\begin{equation}
  V^\mu = V_1^\mu +V_2^\mu+\cdots+ V_n^\mu
  \label{eq:Vmuindep}
\end{equation}
which, because of their independence, must each \emph{separately} satisfy a
gauge-invariance condition similar to (\ref{eq:divVmu}), i.e.,
\begin{equation}
  k_\mu V_i^\mu = V_i \hQ -\hQ V_i~,\qquad i=1,2,\ldots, n~,
  \raisebox{-1.9ex}{}  
  \label{eq:4divVi}
\end{equation}
otherwise the total current $V^\mu$ could not satisfy (\ref{eq:divVmu}).

Let us assume now that one of the $V_i$ is such that the construction
$V^\mu_i=-\{V\}^\mu$ is \emph{not} readily available. In this case, we may
devise an auxiliary phenomenological current $\tV^\mu_i$ instead that satisfies
exactly the same four-divergence relation (\ref{eq:4divVi}) as it would have to
be satisfied by $V^\mu_i$ if it were available. To this end, we adapt the
procedure used successfully in Refs.~\cite{H97,HBMF98,HNK06} for the four-point
interaction currents $M^\mu_\IC$ of meson production to the present
five-point-current case and make the ansatz
\begin{widetext}
\begin{align}
V^\mu_i \to \tilde{V}^\mu_i &=
-\Big[V_i(p_1',p_2';p_1-k,p_2)-W_i(k,p'_1,p'_2;p_1,p_2)\Big]\frac{Q_1(2p_1-k)^\mu}{(p_1-k)^2-p_1^2}
\nonumber\\[2ex]
&\qquad\mbox{}
-\Big[V_i(p_1',p_2';p_1,p_2-k)-W_i(k,p'_1,p'_2;p_1,p_2)\Big]\frac{Q_2(2p_2-k)^\mu}{(p_2-k)^2-p_2^2}
\displaybreak[0]\nonumber\\[2ex]
&\qquad\quad\mbox{}
-\frac{Q_1(2p_1'+k)^\mu}{(p_1'+k)^2-p_1^{\prime\,2}}\Big[V_i(p_1'+k,p_2';p_1,p_2)-W_i(k,p'_1,p'_2;p_1,p_2)\Big]
\nonumber\\[2ex]
&\qquad\qquad\mbox{}
-\frac{Q_2(2p_2'+k)^\mu}{(p_2'+k)^2-p_2^{\prime\,2}}\Big[V_i(p_1',p_2'+k;p_1,p_2)-W_i(k,p'_1,p'_2;p_1,p_2)\Big]~,
\label{eq:GIPcurr1}
\end{align}
where $W_i$ is a function to be chosen to ensure that each term here is free of
propagator singularities. The four-divergence of this auxiliary current is then
readily seen to produce indeed
\begin{equation}
  k_\mu \tV_i^\mu = V_i\hQ -\hQ V_i
\end{equation}
since
\begin{equation}
W_i(k,p'_1,p'_2;p_1,p_2)\left(Q_1+Q_2\right)_{\text{initial}}-\left(Q_1+Q_2\right)_{\text{final}}W_i(k,p'_1,p'_2;p_1,p_2)=0
\end{equation}
vanishes because of charge conservation. The ansatz (\ref{eq:GIPcurr1}) thus
preserves the gauge-invariance condition for the entire interaction. As to how
to choose $W_i$, one of the simplest possibilities is
\begin{align}
W_i(k,p'_1,p'_2;p_1,p_2) &=V_i(p_1',p_2';p_1,p_2)
\nonumber\\
&\mbox{}\quad
+\left[(p_1-k)^2-p_1^2\right]\left[(p_2-k)^2-p_2^2\right]\left[(p_1'+k)^2-p_1^{\prime\,2}\right]\left[(p_2'+k)^2-p_2^{\prime\,2}\right]
                R_i(k,p'_1,p'_2;p_1,p_2)~,
\end{align}
where $R_i(k,p'_1,p'_2;p_1,p_2)$, except for symmetry constraints, is
largely arbitrary (and may be equal to zero). This choice ensures that the
resulting GIP current is free of kinematic singularities, as it must be. In
detail, one thus has
\begin{align}
\tV_i^\mu &=
-\frac{V_i(p_1',p_2';p_1-k,p_2)-V_i(p'_1,p'_2;p_1,p_2)}{(p_1-k)^2-p_1^2}Q_1(2p_1-k)^\mu
-\frac{V_i(p_1',p_2';p_1,p_2-k)-V_i(p'_1,p'_2;p_1,p_2)}{(p_2-k)^2-p_2^2}Q_2(2p_2-k)^\mu
\displaybreak[0]\nonumber\\[2ex]
&\qquad\mbox{}
-Q_1(2p_1'+k)^\mu\frac{V_i(p_1'+k,p_2';p_1,p_2)-V_i(p'_1,p'_2;p_1,p_2)}{(p_1'+k)^2-p_1^{\prime\,2}}
-Q_2(2p_2'+k)^\mu\frac{V_i(p_1',p_2'+k;p_1,p_2)-V_i(p'_1,p'_2;p_1,p_2)}{(p_2'+k)^2-p_2^{\prime\,2}}
\displaybreak[0]\nonumber\\[2ex]
&\qquad\mbox{}
+R_i(k,p_1',p_2';p_1,p_2)\left[(p_1'+k)^2-p_1^{\prime\,2}\right]\left[(p_2'+k)^2-p_2^{\prime\,2}\right]
\nonumber\\[2ex]
&\qquad\quad\mbox{}\times
\Big\{   \left[(p_2-k)^2-p_2^2\right]Q_1(2p_1-k)^\mu+ \left[(p_1-k)^2-p_1^2\right] Q_2(2p_2-k)^\mu\Big\}
\nonumber\\[2ex]
&\qquad\qquad\mbox{}+\Big\{Q_1(2p_1'+k)^\mu \left[(p_2'-k)^2-p_2^{\prime\,2}\right]
 +Q_2(2p_2'+k)^\mu\left[(p_1'-k)^2-p_1^{\prime\,2}\right]\Big\}
\nonumber\\[2ex]
&\qquad\qquad\quad\mbox{}\times
 \left[(p_1-k)^2-p_1^2\right]\left[(p_2-k)^2-p_2^2\right]R_i(k,p_1',p_2';p_1,p_2)
~. \label{eq:GIPcurr2}
\end{align}
\end{widetext}
Note that the subtracted potential contribution $V_i(p'_1,p'_2;p_1,p_2)$ is
unphysical since $p_1'+p_2'\neq p_1+p_2$.

Hence, the procedure just outlined allows a systematic hybrid treatment of all
independent contributions $V_i$ to the full interaction $V$ where some
five-point currents $V_i^\mu$ can be treated explicitly and some may be
replaced by auxiliary currents $\tV^\mu_i$ according to (\ref{eq:GIPcurr1}).
The total interaction current $V^\mu$ will satisfy the condition
(\ref{eq:divVmu}) as a matter of course and gauge invariance is not at issue.

\subsection{\boldmath Extension to coupled channels: $NN$, $N\Delta$, and $\Delta\Delta$}\label{sec:delta}

The derivation of the bremsstrahlung current in Sec.~\ref{sec:4dimformalism} is
completely generic and will remain true regardless of the actual mechanisms
taken into account in the nucleon-nucleon interaction $V$ that drives the
Bethe-Salpeter equation (\ref{eq:NNLS}). The current mechanisms depicted in
Fig.~\ref{fig:Jfigred} assume an interaction based on single-meson exchanges
between nucleons only. Here we briefly discuss the necessary modifications if
such exchanges involve transitions between different baryonic states. We limit
the discussion to transitions between the nucleon $N$ and the $\Delta(1232)$
mediated by single-meson exchanges, i.e., we consider the effect of coupling
the channels $NN$, $N\Delta$, and $\Delta\Delta$; transitions into other
(resonant) baryonic states can be treated along the same lines.

It is a very simple and straightforward exercise to decouple the corresponding
set of Bethe-Salpeter equations that couple the $NN$, $N\Delta$, and
$\Delta\Delta$ channels and write the $NN$ interaction appropriate for the
\emph{single}-channel Bethe-Salpeter equation (\ref{eq:NNLS}) as
\begin{equation}
  V = V_{\textsc{mec}} + V_\Delta~.
 \label{eq:totalV}
\end{equation}
The first term, $V_{\textsc{mec}}$, describes single-meson exchanges
between nucleons, as shown on the left-hand side of
Fig.~\ref{fig:Vmudetail}, that provide the current mechanisms depicted on the
right-hand side of the figure. The second term, $V_\Delta$, contains all
intermediate $N \Delta$ or $\Delta\Delta$ contributions and their transitions.
Using the notations
\begin{subequations}\label{eq:Uexch}
\begin{equation}
  U_N \text{:\quad meson-exchange transition $NN\to N\Delta$}
\end{equation}
and
\begin{equation}
  U_\Delta \text{:\quad meson-exchange transition $NN\to \Delta\Delta$}
\end{equation}
\end{subequations}
for the transition interactions that mediate the coupling to the primary $NN$
channel, we obtain
\begin{align}
  V_\Delta &= U^\dagger_{N} \left(G^{\Delta}_{N}+G^{\Delta}_{N}T^\Delta_{N,N}G^{\Delta}_{N} \right) U_N
  +U^\dagger_{\Delta} G^{\Delta}_{\Delta}T^\Delta_{\Delta,N}G^\Delta_{N} U_N
\nonumber\\
&\mbox{}\quad
  + U^\dagger_\Delta \left( G^\Delta_\Delta +G^\Delta_\Delta T^\Delta_{\Delta,\Delta} G^\Delta_\Delta\right) U_\Delta
  +U^\dagger_{N} G^{\Delta}_N T^\Delta_{N,\Delta}G^\Delta_{\Delta} U_\Delta
  \label{eq:VDelta}
\end{align}
where $G^\Delta_N$ and $G^\Delta_\Delta$ describe the intermediate propagation
of the $\Delta N$ and the $\Delta\Delta$ systems, respectively. Intermediate
transitions $\Delta N \to \Delta N$,  $\Delta N \to \Delta \Delta$, $\Delta
\Delta\to \Delta N$ $\Delta\Delta \to \Delta\Delta$ are subsumed in the
respective $T$ matrices $T^\Delta_{N,N}$, $T^\Delta_{\Delta,N}$,
$T^\Delta_{N,\Delta}$, and $T^\Delta_{\Delta,\Delta}$.

%
\begin{figure}[t!]\centering
  \includegraphics[width=.95\columnwidth,clip=]{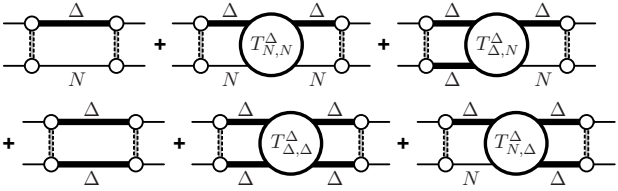}
  \caption{\label{fig:DeltaBoxes}%
  Contributions to the $NN$ interaction involving intermediate $N\Delta$ and
  $\Delta\Delta$ systems and intermediate transitions $T^\Delta_{fi}$ between
  such systems as given by $V_\Delta$ in Eq.~(\ref{eq:VDelta}).
  Summations over all possible exchanged mesons that mediate transitions
  to the $\Delta$ state are implied. Here, we take into account $\pi$ and
  $\rho$ exchanges. Antisymmetrization of external nucleons is implied.}
\end{figure}
%

Figure~\ref{fig:DeltaBoxes} provides a graphical representation of $V_\Delta$.
Coupling a photon to each of the mechanisms depicted here results in the
interaction current
\begin{equation}
V^\mu_\Delta = -\{V_\Delta\}^\mu
\end{equation}
which, together with the interaction current $V^\mu_{\MEC}$ depicted
generically in Fig.~\ref{fig:Vmudetail}, constitutes the total interaction
current
\begin{equation} V^\mu = V^\mu_\MEC + V^\mu_\Delta
 \label{eq:totalVmu}
\end{equation}
(if nucleons and $\Delta$'s are the only baryon degrees of freedom considered).
Of course, calculating $V^\mu_\Delta$ explicitly is a formidable task. There
are eight current contributions for each of the simple box graphs (where the
photon can couple to each of the four intermediate hadrons and the four
vertices) and eleven for each of the graphs involving an intermediate $T$
matrix. Considering only $\pi$ and $\rho$ exchanges for these graphs involving
the $\Delta$ and, allowing for the four possibilities of exchanging these
mesons, altogether, therefore, there are 240 current terms. This is
\emph{without} separately accounting for the internal mechanisms resulting from
coupling the photon to the $T$ matrices with $\Delta$ degrees of freedom which
is similar in complexity to the $NN$ bremsstrahlung current itself. In view of
this complexity, we forego drawing the corresponding diagrams for
$V^\mu_\Delta$.

Each set of currents resulting from one graph in Fig.~\ref{fig:DeltaBoxes}
corresponds to an \emph{independent} interaction current, as discussed in
conjunction with (\ref{eq:Vmuindep}), and therefore must satisfy the
gauge-invariance condition (\ref{eq:4divVi}) independently. If an exact
treatment of the corresponding interaction current is not feasible in view of
the complexity of the problem, from the point of view of gauge invariance this
may also be done in an approximate manner by constructing an auxiliary current
for each of the graphs in Fig.~\ref{fig:DeltaBoxes} along the lines discussed
in the preceding Sec.~\ref{sec:phenIntCurrent}.

%
\begin{figure}[t!]\centering
  \includegraphics[width=.6\columnwidth,clip=]{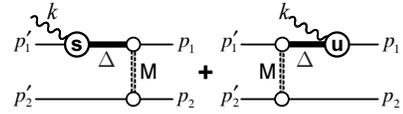}
  \caption{\label{fig:DeltaCurrents}%
  Lowest-order electromagnetic $\gamma N\Delta$-transition contribution to
  $J^\mu_\Delta$  of Eq.~(\ref{eq:JmuDelta}); the exchanged mesons are
  $M=\pi,\rho$. The full contribution involves replacing the single-meson
  exchanges in the two graphs by the $T$-matrix representing the sum of all
  two-nucleon irreducible transitions $NN\to \Delta N$ and $\Delta N\to NN$,
  respectively. The transition currents are transverse and thus have no bearing
  on gauge invariance. Antisymmetrization of nucleons is implied.}
\end{figure}
%

In addition to the hadronic transitions into intermediate $\Delta$ states,
there are also direct electromagnetic transitions $\gamma N\to\Delta$. Such
contributions \emph{cannot} be obtained by applying the gauge-derivative method
used in Sec.~\ref{sec:4dimformalism}, or by any other method based on minimal
substitution. They must be added by hand, in terms of their own Lagrangian [see
Appendix, Eq.~(\ref{VNR32})]. Their full contribution to the basic production
current $J^\mu$ of (\ref{eq:JmuModified}) may be written as
\begin{equation}
  J^\mu_\Delta = d^\mu_{N\Delta} \, G^\Delta_N \, T_{\Delta N, NN}
   + T_{NN,\Delta N}\,G^\Delta_N\, d^\mu_{\Delta N}~,
  \label{eq:JmuDelta}
\end{equation}
where $T_{\Delta N, NN}$ and  $T_{ NN,\Delta N}$ are the $T$-matrices resulting
from summing all two-nucleon irreducible transitions $NN\to \Delta N$ and
$\Delta N\to N N$, respectively; $d^\mu_{N\Delta}$ and $d^\mu_{\Delta N}$, in
an obvious schematic notation borrowed from the $NN$ contributions, contains
the electromagnetic transition current $\Delta\to N$ and $N\to\Delta$,
respectively, along the baryon lines that emit the photon, as depicted in
Fig.~\ref{fig:DeltaCurrents} for the lowest-order single-meson exchange
contributions to the respective $T$-matrices. These contributions are
manifestly transverse and hence have no impact on gauge invariance.
Graphically, they may be subsumed in the right-most diagrams of
Figs.~\ref{fig:Jgeneric} or \ref{fig:Jfigred}, depending on whether one
considers the full formalism or its three-dimensional reduction, respectively.
In other words, $J^\mu_\Delta$ given above is part of the transverse current
$T^\mu$, as anticipated already in Eq.~(\ref{eq:TmuStructure}).

In summary, the full basic interaction current for $NN$ bremsstrahlung,
including $\Delta$ degrees of freedom, becomes
\begin{equation}
   J^\mu = d^\mu G_0 V + V G_0d^\mu + V^\mu
           + \left[ J^\mu_M + J^\mu_{\Delta} - VG_0d^\mu G_0V \right] ~,
\end{equation}
where the groupings of the four terms corresponds to the four graphs on the
right-hand side of Fig.~\ref{fig:Jgeneric}. Here, $V$ and $V^\mu$ contain
$\Delta$ degrees of freedom according to Eqs.~(\ref{eq:totalV}) and
(\ref{eq:totalVmu}), respectively, and the transverse current is taken as
$T^\mu=J^\mu_M + J^\mu_{\Delta}$. In the reduced case, we have
 \begin{equation}
 J^\mu_r =
d^\mu G_0 V + VG_0d^\mu
 +V^\mu
 + \left[ J^\mu_M +J^\mu_{r\Delta} \right]~,
            \label{eq:JmurMD}
\end{equation}
where the last grouping is the explicit transverse current $\JT^\mu=J^\mu_M
+J^\mu_{r\Delta}$, with $\lambda_i$ and $\lambda_f$ of Eq.~(\ref{eq:bmuT1}) put
to zero; the reduced $N\Delta$-transition current $J^\mu_{r\Delta}$ is obtained
by the corresponding three-dimensional reductions of the loop integrations
within $J^\mu_{\Delta}$.

In our present results (discussed in the next section), in view of their
complexity, we do not take into account any currents that account for the
box-graph mechanisms of Fig.~\ref{fig:DeltaBoxes}, i.e., $V^\mu_\Delta$ is set
to zero. However, we do take into account the transition-current contributions
(\ref{eq:JmuDelta}) in lowest order, as depicted in
Fig.~\ref{fig:DeltaCurrents}.

\section{Applications}\label{sec:Application}

The present formalism was applied successfully~\cite{NH09} in the description
of the high-precision KVI data~\cite{KVI02} at a proton incident energy of 190
MeV. In this case, in view of the relatively low energy, the dominant current
contribution is expected to be essentially nucleonic. We felt justified,
therefore, to restrict ourselves to the $NN$ channel only and neglect the
effects of coupling to other baryon-baryon channels. Here, we apply the
formalism to some selected data sets from TRIUMF \cite{TRIUMF90} at a higher
incident energy of 280 MeV. This energy is just above the pion-production
threshold energy and still well below the $\Delta(1232)$ resonance peak energy
region of about 650 MeV. Nevertheless, an earlier analysis of
Ref.~\cite{dJNL95} within a coupled-channel approach (employing $NN$ and
$N\Delta$ channels) indicated that incorporating $\Delta$ contributions leads
to better agreement with the data at certain geometries. For the TRIUMF
experiment at 280 MeV, in particular, the dominant $\Delta$ contribution was
found to be arising from the $\gamma N\Delta$ transition current,
$J^\mu_\Delta$, in lowest order \cite{dJNL95} (a similar result was found in
the single-channel approach of Ref.~\cite{dJNHS94}). We point out, however,
that the approach reported in Ref.~\cite{dJNL95} is not gauge invariant beyond
the soft photon approximation. Since here we wish to preserve full gauge
invariance, we need to follow the procedure outlined in Sec.~\ref{sec:delta}
for the incorporation of $\Delta$ degrees of freedom. In general, however, this
is too challenging technically at present, \emph{except} for the lowest order
of the transition-current contributions of Eq.~(\ref{eq:JmuDelta}) depicted in
Fig.~\ref{fig:DeltaCurrents} which we do take into account in the results
presented below. Incorporation of the interaction current $V^\mu_\Delta$
arising from the $\Delta$-box-diagram contributions of
Fig.~\ref{fig:DeltaBoxes} is beyond the scope of the present paper and we will
leave this for future studies.

The details of the present model for constructing the production current
$J^\mu$ as well as the $NN$ interaction are given in the Appendix.

Figure~\ref{fig:dxsc} shows a comparison of the present results for the cross
sections with the TRIUMF data in the coplanar geometry at small proton
scattering angles of $\theta_1=12^\circ$ and $\theta_2=14^\circ$ and at larger
proton scattering angles of $\theta_1=28^\circ$ and $\theta_2=27.8^\circ$. We
mention that in the present work, we follow the convention of
Ref.~\cite{KVI02,KVI04} for assigning the nucleon angles $\theta_1$ and
$\theta_2$ which is just opposite to the one adopted in other studies
\cite{HNSA95,dJNHS94,dJNL95}.  Also, no arbitrary normalization factor of 2/3
to the data points \cite{TRIUMF90} is applied here. The parameters $h$ and $a$
of the phenomenological four-point contact current determined in
Ref.~\cite{NH09} in the description of the KVI data are kept at the same
values, $h=2.5$ and $a=1000$, despite the fact that they had been determined at
the much lower incident energy of 190 MeV. In principle, of course, these
parameters may be functions of the incident energy. We see that the agreement
is very reasonable. The influence of the $\Delta$-resonance contribution is
relatively small but significant and definitely helps to improve the agreement
with the data. This is in line with the earlier findings of
Refs.~\cite{dJNHS94,dJNL95}. Furthermore, the present results are of comparable
quality to the earlier results \cite{dJNHS94,dJNL95} in reproducing the TRIUMF
data, indicating that, unlike for the high-precision KVI data at small proton
scattering angles \cite{KVI02,KVI04}, the generalized four-point contact
current --- required to maintain the gauge invariance of the full
bremsstrahlung amplitude --- is apparently not that critical. The dotted curves
in Fig.~\ref{fig:dxsc} correspond to the results when the generalized
four-point contact current is switched off. As one can see, its effect is
visible but somewhat smaller than the effect of the $\Delta$ contribution.
Overall, however, apart from the differences in reaction energies and
scattering-angle configurations, we note that the TRIUMF data are much less
precise than the KVI data, thus making the assessment of how relevant various
theoretical contributions are in reproducing the data much less conclusive.

%
\begin{figure}[t!]\centering
\includegraphics[width=.9\columnwidth,clip=]{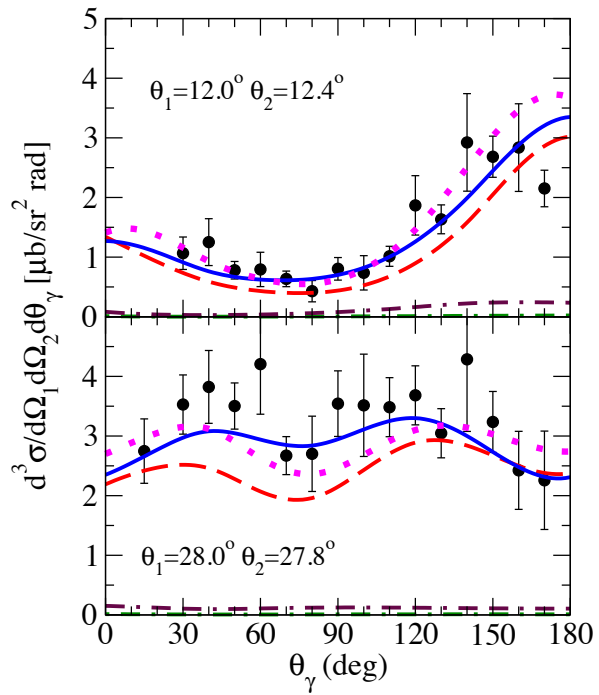}
\caption{\label{fig:dxsc}%
(Color online) Comparison of the present prediction for the cross sections with
the TRIUMF data \cite{TRIUMF90} in coplanar geometry at 280-MeV proton incident
energy. $\theta_1$ and $\theta_2$ denote the fixed scattering angles of the two
protons in the final state. The cross sections are shown as functions of the
emitted photon angle, $\theta_\gamma$, in the laboratory frame. The
(blue) solid curves show the results with the total current $J^\mu_r$ of
Eq.~(\ref{eq:JmurMD}). The (red) dashed curves correspond to the results when
the transition current, $T^\mu_r=J^\mu_M+J^\mu_{r \Delta}$, is switched off in
its entirety. The transition-current contributions are shown separately as
(green) dash-dotted curves (which are hardly visible in the graphs) for the
meson transitions ($\gamma\pi\rho$ and $\gamma\pi\omega$) subsumed in
$J^\mu_M$ and as (brown) dot-double-dashed curves for the $\gamma N\Delta$
transitions that provide the lowest-order contributions of $J^\mu_{r \Delta}$.
The (magenta) dotted curves correspond to the results when the generalized
four-point contact current (cf. Fig.~\ref{fig:Jfigred}) is switched off.
}
\end{figure}
%

%
\begin{figure}[t!]\centering
\includegraphics[width=0.9\columnwidth,clip=]{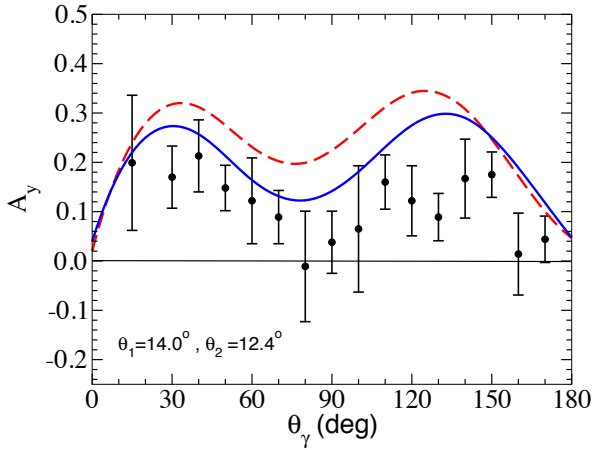}
\caption{\label{fig:Ay}%
(Color online) Comparison of the present prediction for the analyzing power
with the TRIUMF data \cite{TRIUMF90} in coplanar geometry at 280-MeV proton
incident energy. $\theta_1$ and $\theta_2$ denote the fixed scattering angles
of the two protons in the final state. The analyzing power is shown as a
function of the emitted photon angle, $\theta_\gamma$, in the laboratory frame.
The (blue) solid line is obtained with the total current $J^\mu_r$ of
Eq.~(\ref{eq:JmurMD}). The (red) dashed curve corresponds to the result when
the $\gamma N\Delta$ transition current, $J^\mu_{r \Delta}$, is switched off.}
\end{figure}
%

In Fig.~\ref{fig:Ay}, we show our results for the analyzing power together with
the TRIUMF data at $\theta_1=12^\circ$ and $\theta_2=12.4^\circ$. The agreement
is reasonable given the uncertainties in the data. Again, the
$\Delta$-resonance current helps improve the agreement with the data.

In summary, therefore, our results for the TRIUMF data at a proton incident
energy of 280 MeV are in agreement with the findings of the previous studies
\cite{dJNHS94,dJNL95}. We point out, however, that our results are obtained
within a fully gauge-invariant framework. Moreover, one may hope that inclusion
of the $\Delta$-box contributions may further improve the agreement.

\section{Summary}\label{sec:Summary}

We have presented a complete, rigorous formulation of the $NN$ bremsstrahlung
reaction based on a relativistic field-theory approach in which the photon is
coupled in all possible ways to the underlying two-nucleon $T$-matrix obtained
from the corresponding covariant Bethe-Salpeter-type $NN$ scattering equation
using the gauge-derivative procedure of Haberzettl~\cite{H97}. The
resulting bremsstrahlung amplitude is unitary as a matter of course and it
satisfies full local gauge invariance as dictated by the generalized
Ward-Takahashi identity. The novel feature of this approach is the consistent
--- i.e., gauge-invariant --- incorporation of interaction currents resulting
from the photon coupling internally to interacting hadronic systems.

The formalism is quite readily adapted to approximations and thus can be
applied even in cases where the microscopic dynamical structure of the
underlying interacting hadronic systems is either not known in detail or too
complex to be treated in detail. We have pointed out how the interaction
currents resulting from the photon being attached to nucleon-nucleon-meson
vertices can be treated by phenomenological four-point contact currents that
preserve gauge invariance following the approach of Haberzettl, Nakayama, and
Krewald \cite{HNK06}. In an advance application of the present formalism
\cite{NH09}, such interaction currents had been shown to contribute
significantly to reproducing the high-precision proton-proton bremsstrahlung
data at 190 MeV obtained at KVI \cite{KVI02}, thus removing a longstanding
discrepancy between theory and experiment. In addition, we have provided a
scheme that permits the approximate treatment of current contributions
resulting from pieces of the $NN$ interaction that cannot be incorporated
exactly. In each case, the approximation procedure ensures gauge invariance of
the entire bremsstrahlung amplitude.

We have also discussed the necessary modifications when taking into account
baryonic states other than the nucleon $N$; in detail, we consider the
$\Delta(1232)$ resonance by incorporating the couplings of the $NN$ to the
$N\Delta$ and $\Delta\Delta$ systems, and the $\gamma \to N \Delta$
transitions.

The formalism has been applied to the 280-MeV bremsstrahlung data from TRIUMF
\cite{TRIUMF90} and we find good agreement with the data similar to what was
obtained by other authors. We emphasize, however, that our results obey full
gauge invariance, whereas previous approaches provide a conserved current only
in the soft-photon approximation. Since the gauge-invariant incorporation of
intermediate $N\Delta$ and $\Delta\Delta$ configurations is too demanding
technically at present, the only $\Delta$ degrees of freedom employed in this
calculation have been the $\gamma N\Delta$-transition currents.

Finally, we emphasize that despite its completeness and generality, the present
approach is quite flexible and amenable to approximations, as discussed in
Secs.~\ref{subsec:singlemeson} and \ref{sec:phenIntCurrent}. We expect,
therefore, that it will also be useful in the on-going investigation of hard
bremsstrahlung \cite{ANKE,JW09}, as well as in di-lepton production processes
\cite{HADES} where the photon is virtual.

\appendix*
\section{}\label{app:appendix}

For the present application, we construct the basic production current $J^\mu$
in the framework of a covariant three-dimensional Blankenbecler-Sugar
reduction~\cite{BlankenbeclerSugar} (see also Ref.~\cite{E74}), as described in
Ref.~\cite{NH09}, consistent with the $NN$ FSI and ISI based on the OBEP-B
version of the Bonn potential \cite{Bonn_NN}. In addition to the resulting
production current $J^\mu_r$ used already in Ref.~\cite{NH09}, the present work
includes the $\Delta$-resonance current interaction calculated in terms of the
following Lagrangians:
\begin{subequations}
\begin{align}
\mathcal{L}_{\Delta N\pi} &=
\frac{g_{\Delta N\pi}^{}}{M_\pi} {\bar{\Delta}^\mu} \vec{T}^\dagger\cdot (\partial_\mu \vec\pi) N + \hc \ ,
 \label{PNR32}
\displaybreak[0]\\
\mathcal{L}_{\Delta N\rho} &=
- i\frac{g^{(1)}_{\Delta N\rho}}{2M_N} {\bar{\Delta}^\mu} \gamma^\nu \gamma_5 \vec{T}^\dagger\cdot \vec\rho_{\mu\nu} N
\nonumber\displaybreak[0] \\
&\mbox{}\quad
+ \frac{g^{(2)}_{\Delta N\rho}}{4M^2_N} \bar{\Delta}^\mu \gamma_5 \vec{T}^\dagger \cdot\vec\rho_{\mu\nu} \partial^\nu N
\nonumber\displaybreak[0] \\
&\mbox{}\quad
 - \frac{g^{(3)}_{\Delta N\rho}}{4M^2_N} \bar{\Delta}^\mu \gamma_5 \vec{T}^\dagger\cdot\left(\partial^\nu \vec\rho_{\mu\nu}\right) N  +  \hc \ ,
\displaybreak[0] \\
\mathcal{L}_{\Delta N\gamma} &=
- ie\frac{g^{(1)}_{\Delta N\gamma}}{2M_N} \bar{\Delta}^\mu \gamma^\nu \gamma_5 T^\dagger_3 F_{\mu\nu} N
\nonumber\displaybreak[0] \\
&\mbox{}\quad
 +e \frac{g^{(2)}_{\Delta N\gamma}}{4M^2_N} \bar{\Delta}^\mu \gamma_5 T^\dagger_3 F_{\mu\nu} \partial^\nu N
 + \hc~,
\label{VNR32}
\end{align}
\end{subequations}
where $\vec\rho_{\mu\nu} \equiv \partial_\mu \vec\rho_\nu - \partial_\nu
\vec\rho_\mu$ and $F_{\mu\nu} \equiv \partial_\mu A_\nu - \partial_\nu A_\mu$.
The coupling constants are taken to be $g_{\Delta N\pi}=2.12$ and
$g^{(1)}_{\Delta N\rho}=40.5$ \cite{Bonn_NNfull}; for simplicity,
$g^{(2)}_{\Delta N\rho}$ and $g^{(3)}_{\Delta N\rho}$ are set to zero. The
electromagnetic coupling constants $g^{(1)}_{\Delta N\gamma}=5.00$ and
$g^{(2)}_{\Delta N\gamma}=-4.73$ are extracted from the corresponding helicity
amplitudes \cite{PDG}. The $NN\pi$ and $NN\rho$ vertices are derived from the
corresponding Lagrangians used for constructing the nucleonic current in
Ref.~\cite{NH09}.

Also for simplicity, the $\Delta N\pi$ and $\Delta N\rho$ vertices are provided
with the same form factors as in the $NN\pi$ and $NN\rho$ vertices \cite{NH09}
for the off-shell meson and baryon.

For the $\Delta$ propagator, we use
\begin{align}
S^{\mu\nu}(p) & = \frac{\fs{p} + M_\Delta}{p^2 - M_\Delta + i\frac{\Gamma_\Delta}{2}}
 \nonumber \\
 &\quad\mbox{} \times \bigg( -g^{\mu\nu}  -  \frac{1}{3M_\Delta}(p^\mu\gamma^\nu - p^\nu\gamma^\mu)
  \nonumber \\
 &\mbox{}\qquad\qquad
   + \frac{1}{3}\gamma^\mu\gamma^\nu   + \frac{2}{3M_\Delta^2} p^\mu p^\nu \bigg) \ ,
\label{eq:D-prop}
\end{align}
where $M_\Delta = 1232$ MeV and $\Gamma_\Delta=120$ MeV.

\acknowledgments
This work is supported in part by the FFE-COSY Grant No.\
41788390.


\end{document}